\documentclass[12pt]{article}
\begin{document}
\input amssym.tex 

\title{Remarks on a five-dimensional Kaluza-Klein theory of the massive 
Dirac monopole}

\author{Ion I.  Cot\u aescu \thanks{E-mail:~~~cota@physics. uvt. ro}\\ 
{\small \it West University of Timi\c soara,}\\
       {\small \it V.  P\^ arvan Ave.  4, RO-300223 Timi\c soara, Romania}}


\maketitle

\begin{abstract}

The Gross-Perry-Sorkin spacetime, formed by the Euclidean Taub-NUT space 
with the time trivially added, is the appropriate background of the Dirac 
magnetic monopole without an explicit mass term. One remarks that there 
exists a very simple five-dimensional metric of spacetimes carrying massive 
magnetic monopoles that is an exact solution of the vacuum Einstein equations. 
Moreover, the same isometry properties as the original Euclidean Taub-NUT 
space are preserved. This leads to an Abelian Kaluza-Klein theory whose metric 
appears as a combinations between the Gross-Perry-Sorkin and Schwarzschild ones. The 
asymptotic motion of the scalar charged test particles is discussed, now by 
accounting for the mixing between the gravitational and magnetic effects.       

Pacs 04. 62. +v

\end{abstract}

\newpage

A special class of solutions of the Maxwell or Yang-Mills equations are the 
instantons and monopoles defined on  appropriate flat or curved backgrounds
\cite{AT,EGH,AT1}. A natural framework is offered by the Kaluza-Klein theories
where the gauge degrees of freedom deal with specific extra-coordinates 
exceeding the physical spacetime. In these theories the basic problem is to 
find the solutions of the field equations in geometries whose global metrics 
should be  exact solutions of the Einstein equations.

A typical example is the four-dimensional Euclidean Taub-NUT space which
involves the potentials of the Dirac magnetic  monopole \cite{D} and
satisfies the vacuum Einstein equations \cite{TNUT}. Moreover, this space 
is hiper-K\" ahler having many interesting properties related to a 
specific hidden symmetry \cite{GM,NOVA}. For this reason the K\" ahlerian 
geometries were considered for generalizing the Dirac monopole to many 
dimensions \cite{ABE}. Other successful generalizations were obtained 
integrating the field equations in Kaluza-Klein theories with five or more 
dimensions \cite{56N,SGPS}       

In this way a large collection of metrics was found including some metrics 
corresponding to massive Dirac monopoles with explicit mass terms. Thus  
the whole set of the monopole metrics can be divided in metrics of the 
Schwarzschild or Gross-Perry-Sorkin types \cite{SGPS}. In this letter we  
show that there exists a hybrid metric giving rise to a simple Abelian 
five-dimensional Kaluza-Klein theory of a monopole with gravitational mass.  
This metric is an asymptotic flat solution of the time-dependent vacuum 
Einstein equations combining Schwarzschild terms with Gross-Perry-Sorkin ones. 
Our purpose is to evaluate the mixing between the magnetic and gravitational 
effects produced by this metric in the asymptotic domain.

The Euclidean Taub-NUT manifold is the space  of the Abelian Kaluza-Klein
theory of the Dirac magnetic monopole that  provides a non-trivial 
generalization of the Kepler problem.  This space is a static four-dimensional 
manifold, ${ M}_4\sim {\Bbb R}^4$, equipped with the isometry group 
$SO(3)\otimes U(1)$ and carrying the the Dirac magnetic 
monopole related to  strings along the third axis.  
This symmetry recommends to use local charts with spherical 
coordinates $(r,\theta, \varphi, {\alpha})$ where the first three are the 
usual spherical coordinates of the vector $\vec{x}=(x^1,x^2,x^3)$, with 
$|\vec{x}|=r$, while ${\alpha}$ is the angular Kaluza-Klein extra-coordinate.  
The spherical coordinates can be associated with the Cartesian ones 
$(\vec{x},y)=(x^1,x^2,x^3,y)$ where the extra-coordinate $y$ may depend linearly  
on $\alpha$.  In Cartesian coordinates one has the opportunity 
to use the vector notation and the scalar products 
$\vec{x}\cdot \vec{x}^{\prime}$ which are invariant under the $SO(3)$ rotations.

The Euclidean Taub-NUT space  has the virtue to be Ricci flat, its metric 
being an exact solution of the vacuum Einstein equations \cite{TNUT}. 
In the Cartesian charts $(\vec{x},y^{\pm})$ the line elements read
\begin{equation}\label{dsTNUT}
{ds_{\pm}}^2=G(r)\,d\vec{x}\cdot d\vec{x}+
G(r)^{-1}(dy^{\pm}+\vec{A}^{\pm}\cdot d\vec{x})^2\,, 
\end{equation}
where 
\begin{equation}
G(r)=1+\frac{\mu}{r}\,.
\end{equation}
The vector potential of the Dirac magnetic monopole produced by a string along 
the negative third semi-axis is denoted by $\vec{A}^+$ while $\vec{A}^{-}$ is 
due to a string along the positive one. These potentials have the components  
\begin{equation}\label{pot}
A_1^{\pm}=\mp \frac{\mu\,x^2}{r(r\pm x^3)}\,, \quad
A_2^{\pm}=\pm \frac{\mu\,x^1}{r(r\pm x^3)}\,, \quad
A_3^{\pm}=0\,. 
\end{equation}
The potential $\vec{A}^-$  differs from $\vec{A}^+$ only within a gauge, 
giving rise to the same magnetic field with central symmetry,
\begin{equation}
\vec{B}={\rm rot}\vec{A}^{\pm}=\mu\, \frac{\vec{x}}{r^3}\,. 
\end{equation}
In this manner the original string singularity is reduced to a point-like 
one interpreted as a magnetic monopole with the magnetic charge $\mu$. 

The crucial point of this construction is the correct definition of the 
transition function between the Cartesian charts $(\vec{x},y^{\pm})$ or the 
corresponding spherical ones $(r,\theta,\varphi,\alpha^{\pm})$ \cite{EGH}. 
It is obvious that the transition $y^{-}=y^{+}+2\mu\varphi$ is in accordance 
with the gauge transformation $\vec{A}^-\cdot d\vec{x}=\vec{A}^+\cdot d\vec{x}-
2\mu d\varphi$. This transition defines a suitable non-trivial fibration which 
is a version of the Hopf one, $S^3\to S^2$ \cite{EGH}. In these conditions it 
is convenient to take $y^{\pm}=-\mu(\alpha^{\pm}\pm\varphi)$ which leads to 
$\alpha^-=\alpha^+=\alpha$. Then both the line elements (\ref{dsTNUT}) get the 
same form, 
\begin{equation}
{ds_o}^2=G(r)\,(dr^2+r^2d\theta^2 + r^2 \sin^2\theta d\varphi^2)
+\mu^2 G(r)^{-1}(d\alpha+ \cos\theta d\varphi)^2\,, 
\end{equation}
in the spherical charts where the vector potentials have the components
\begin{equation}\label{AA}
A_r^{\pm}=A_{\theta}^{\pm}=0\,, \quad A^{\pm}_{\varphi}=\mu(\pm 1-\cos\theta)\,.
\end{equation}

In order to implement the Euclidean Taub-NUT space in physical Kaluza-Klein 
theories one has to include the time $t$. The Gross-Perry-Sorkin metric, 
\begin{equation}\label{trt}
{d\hat s}^2=-dt^2+{ds_{o}}^2\,, 
\end{equation}
has the remarkable property to remain  Ricci flat \cite{TNUT} since it is a 
solution of the vacuum Einstein equations in five dimensions. Moreover, this 
leads to a geometry with $SO(3)\otimes U(1)\otimes T(1)_t$ isometries where 
$T(1)_t$ is the group of the time translations. The metric (\ref{trt}) 
incorporates the effects of the magnetic charge $\mu$ while the gravitational 
interaction seems to be rather a consequence of the magnetic one. The reason 
is that there are no terms of the Schwarzschild type with at least one 
parameter which could be interpreted as the gravitational mass of the magnetic 
monopole.   

Under such circumstances we assume that the metric (\ref{trt}) describes  
massless monopoles, but we have to look for another simple five-dimensional 
Kaluza-Klein metric suitable for massive Dirac monopoles.  In these geometries 
we maintain the $SO(3)\otimes U(1)\otimes T(1)_t$ isometries as well as the 
form of the potentials (\ref{pot}), up to the factor representing the 
magnetic charge. Requiring the whole spacetime to be Ricci flat, we find an 
interesting solution of the vacuum Einstein equations with the line element,
\begin{eqnarray}\label{ds}
ds^2&=&-F(r)dt^2+G(r)\left(\frac{dr^2}{F(r)} +r^2 d\theta^2+
r^2\sin^2\theta d\varphi^2\right)\nonumber\\
&&+\,\frac{\mu_{ef}^2 }{G(r)}\,(d\alpha +\cos\theta d\varphi)^2\,, 
\end{eqnarray}
where both $\mu_{ef}=\sqrt{\mu(\mu +2M)}$ and 
\begin{equation}
F(r)=1-\frac{2M}{r}
\end{equation}
depend on the monopole mass $M$. This metric is asymptotic flat embedding 
gravitational and magnetic terms. We observe that for $M\to 0$ we recover the 
line element (\ref{trt}) while for $\mu\to 0$ the magnetic properties disappear. 
In the latter case the metric reduces  to the usual Schwarzschild one of a 
particle with the mass $M$. It is remarkable that the massive monopole has 
the {\em effective} magnetic charge $\mu_{ef}$, which depends on the genuine 
magnetic charge $\mu$ and the monopole mass $M$. Obviously, the non-vanishing 
components of the potentials (\ref{AA}) become now 
$A_{\varphi}^{\pm}=\mu_{ef}(\pm 1 -\cos\theta)$.

Next let us consider the motion of a test particle in the gravitational 
and magnetic fields of a massive Dirac monopole at large distances.  
We suppose that the test particle is a quantum scalar particle of the bare 
mass $m$ and charge $e=q/\mu_{ef}$ where $q$ is the eigenvalue of the 
operator $Q=-i\partial_{\alpha}$ \cite{NOVA}. The scalar field $\Phi$ of the 
test particle obeys the five-dimensional Klein-Gordon equation 
$(\square-m^2)\Phi=0$ \cite{BD}. Because of the central symmetry, there are 
particular solutions with separated spherical variables like  
\begin{equation}
\Phi(r,\theta,\phi,\alpha)=R_{E,q,l}(r)Y_{lm}^q(\theta,\phi,\alpha)     
\end{equation}
where $Y_{lm}^q$ are the $SO(3)\otimes U(1)$ harmonics we introduced before 
in Ref. \cite{Y}. The radial functions $R_{E,q,l}$ depend on the energy $E$, 
the angular quantum number $l$, and $q$. Such functions  satisfy the radial 
equation
\begin{eqnarray}
&&\left[-\frac{1}{r^2}\frac{d}{dr}\left(r^2 F\frac{d}{dr}\right)
+\frac{l(l+1)}{r^2}\right.\nonumber\\
&&~~~~~~~~~~~~~~~~\left.+e^2\left(G^2-\frac{\mu_{ef}^2}{r^2}\right)
+m^2 G-E^2\frac{G}{F}\right]R_{E,q,l}(r)=0\,,
\end{eqnarray}
but this can not be analytically solved.

However, for large distances, like $r\gg 2M$, the above equation can be 
approximated by the asymptotic radial equation, 
\begin{equation}\label{rad}
\left[-\frac{d^2}{dr^2}+\frac{l(l+1)}{r^2}-\frac{\nu}{r}\right]
\hat R_{E,q,l}(r)=(E^2 -m_{ef}^2) \hat R_{E,q,l}(r),
\end{equation}
where $\hat R_{E,q,l}(r) \propto r \,R^{asympt.}_{E,q,l}(r)$. The new parameter
\begin{equation}
\nu=2Mm_{ef}^2+(E^2-m_{ef}^2)(\mu+2M)-\mu e^2
\end{equation}
depends on the effective mass of the test particles, $m_{ef}=\sqrt{m^2+e^2}$, 
which includes  the standard  Kaluza-Klein contribution. This time the radial 
equation (\ref{rad}) is analytically solvable. Indeed, one deals with 
a Keplerian motion under the relativistic potential $\phi=-\nu/r$ corresponding 
to the nonrelativistic one $\phi/2m_{ef}$. Moreover, one finds that for low 
energies, like $E\sim m_{ef}$, the 
nonrelativistic potential appears as the Newtonian potential produced by the 
effective monopole mass
\begin{equation}
M_{ef}=M-\frac{\mu}{2}\,\frac{e^2}{m^2+e^2}\,.
\end{equation}

The conclusion is that the metric (\ref{ds}) of the background of a massive 
Dirac monopole mixes the gravitational and the magnetic effects in such a 
manner that for large distances and low energies the gravity may screen the 
magnetism. This could explain why it is so difficult to find experimental 
evidences about possible cosmic objects with magnetic charges.

\subsection*{Acknowledgments}

I should like to thank Erhardt Papp and Mihai Visinescu for interesting and 
useful discussions. This work is partially supported by MEC-AEROSPATIAL Program, Romania.


\end{document}